\documentclass[aps,prl,preprint,psfig,groupedaddress,letterpaper]{revtex4}
\usepackage{amsmath,amssymb}
\usepackage{graphicx}
\usepackage{bm}
\usepackage{epstopdf} 
\usepackage{epsfig}
\usepackage[final]{pdfpages}

\AppendGraphicsExtensions{.tif}

\newcommand{\GMA}{(Ga,Mn)As}
\newcommand{\GMxA}{Ga$_{1-x}$Mn$_x$As}
\newcommand{\GMnA}{Ga$_{0.95}$Mn$_{0.05}$As}
\newcommand{\BSTS}{(Bi,Sb)$_2$(Te,Se)$_3$}

\newcommand{\BS}{Bi$_2$Se$_3$}

\newcommand{\BTSx}{Bi$_2$(Te$_{3-x}$Se$_x$)}
\newcommand{\BTTS}{Bi$_2$Te$_2$Se}
\newcommand{\TC}{$T_{\rm{C}}$}

\renewcommand{\vec}[1]{\mbox{\boldmath$1$}}

\newcounter{lastnote}

\def\bc{\begin{center}}
\def\ec{\end{center}}
\def\be{\begin{equation}}
\def\ee{\end{equation}}
\renewcommand{\vec}[1]{\mbox{\boldmath$1$}}

\begin{document}
\title{Engineering the breaking of time-reversal symmetry in gate-tunable hybrid ferromagnet/topological insulator heterostructures}
\author{Joon Sue Lee}
\thanks{Current address$:$ California NanoSystems Institute, University of California, Santa Barbara, CA 93106}
\email{joonsuelee@engineering.ucsb.edu}
\affiliation{Department of Physics and Materials Research Institute, The Pennsylvania State University, University Park, Pennsylvania 16802, USA}

\author{Anthony Richardella, Robert D. Fraleigh, Chao-xing Liu, Weiwei Zhao}
\affiliation{Department of Physics and Materials Research Institute, The Pennsylvania State University, University Park, Pennsylvania 16802, USA}

\author{Nitin Samarth}
\email{nsamarth@psu.edu}
\affiliation{Department of Physics and Materials Research Institute, The Pennsylvania State University, University Park, Pennsylvania 16802, USA}

\date{\today}
\begin{abstract}
Studying the influence of breaking time-reversal symmetry on topological insulator surface states is an important problem of current interest in condensed matter physics and could provide a route toward proof-of-concept spintronic devices that exploit spin-textured surface states. Here, we develop a new model system for studying the effect of breaking time-reversal symmetry: a hybrid heterostructure wherein a ferromagnetic semiconductor \GMxA{}, with an out-of-plane component of magnetization, is cleanly interfaced with a three-dimensional topological insulator \BSTS{} by molecular beam epitaxy. Lateral electrical transport in this bilayer is dominated by conduction through the topological insulator whose conductivity is a few orders of magnitude higher than that of the highly resistive ferromagnetic semiconductor with a low Mn concentration. Electrical transport measurements of a top-gated heterostructure device reveal a crossover from weak anti-localization (negative magneto-conductance) to weak localization (positive magneto-conductance) as the temperature is lowered or as the chemical potential approaches the Dirac point. This is accompanied by a systematic emergence of an anomalous Hall effect. These results are interpreted in terms of the opening of a gap at the Dirac point as a result of the exchange coupling between the topological insulator surface state and the ferromagnetic ordering in the \GMxA{} layer. Our study shows that this hybrid system is well suited to explore topological quantum phenomena and to realize proof-of-concept demonstrations of topological spintronic devices at cryogenic temperatures.
\end{abstract}
\maketitle

A three-dimensional (3D) topological insulator (TI) is characterized by its surface states which are protected by time-reversal (TR) symmetry\cite{Moore2010,Hasan2010,Qi2010,Fu2007}. The TR symmetry can be broken by doping a TI with magnetic atoms or interfacing a TI surface with a magnetic layer, causing an energy gap opening at the Dirac point\cite{Qi2008,Liu2009,Chen2010,Wray2011}. Unique quantum phenomena resulting from the broken TR symmetry have been proposed: such as a topological magneto-electric effect\cite{Qi2008}, an image magnetic monopole effect\cite{Qi2009}, topological Kerr and Faraday rotation\cite{Qi2008}, and the quantum anomalous Hall effect\cite{Yu2010}. Some of these phenomena have been demonstrated experimentally\cite{Chang2013QAHE,Wu2016}. With a motivation for studying such effects, synthesis and characterization of magnetically doped 3D TIs with transition metals have been studied\cite{Chen2010,Checkelsky2012,Zhang2013,Lee2014,Zhang2012,Choi2011,Song2012}. Angle-resolved photoemission spectroscopy (ARPES) has suggested evidence a gap opening by breaking TR symmetry in magnetically doped TI systems,\cite{Chen2010,Wray2011,Zhang2012,Chang2014} although recent studies point out an alternative mechanism for the gap seen in such studies \cite{Sanchez-Barriga2016}. In addition, a spin-resolved ARPES experiment revealed the hedgehog-like spin textures in the modified surface state of \BS{} films by Mn doping\cite{Xu2012}. 

In this paper, we focus on another way of breaking TR symmetry in the TI surface states, interfacing a TI surface to a ferromagnetic insulator (FMI) with perpendicular magnetization to evidence the broken TR symmetry and ultimately to realize topological quantum phenomena and potential spintronic applications\cite{Pesin2012,Mellnik2014,Deorani2014,Jamali2015,Li2014,Lee2015}. The key advantage of a TI/FMI heterostructure over the magnetically doped TI system is the selective modification of one surface by an adjacent FMI. Magnetic proximity affects only the interfaced surface, and thus magnetic properties or the resulting effects are free from the magnetism of bulk or another surface of the TI layer. So far, a few TI/FMI heterostructures have been experimentally reported using FMIs interfaced with TIs where the chemical potential is located in or near the conduction band\cite{Wei2013,Yang2013,Kandala2013,Lang2013,Jiang2016,Katmis2016,Wang2016}. We note that all these heterostructures involve ferromagnets whose magnetization is in plane. To further separate the FMI-interfaced TI surface from the electrical coupling to the bulk or another surface, the chemical potential needs to be placed in the bulk band gap. Also, to avoid the surface-to-surface tunneling, a TI film needs to be thicker than the critical thickness for hybridization\cite{Neupane2014}. To study the magnetic proximity, a clean, well-defined TI/FMI interface is necessary, and the FMI should have a magnetization component perpendicular to the TI surface to break the TR symmetry.

We demonstrate a new approach for a TI/FMI heterostructure using a dilute magnetic semiconductor \GMA. The ferromagnetic Curie temperature (\TC) and resistivity as well as the magnetic easy axis of \GMA{} films can be engineered by the amount of Mn-doping, annealing, and strain\cite{MacDonald2005,Ku2003,Matsukura2004}. Here, highly resistive \GMxA{} with an out-of-plane magnetization is desired. High resistivity was achieved by using a low Mn-doping of x $\approx$ 0.05 and a perpendicular component of magnetization by growing the \GMA{} film (15 nm) on an InP (111)A substrate by MBE. (See supplementary information for magnetic and structural characterizations.)
An advantage of using \GMA{} for the TI/FMI heterostructure is the well-defined interface, without an amorphous interfacial layer or secondary phases, as was demonstrated for epitaxial growth of Bi-chalcogenide TIs on GaAs (111)\cite{Richardella2010}. After growing the \GMA{} film in a ultrahigh vacuum chamber (low $10^{-10}$ Torr), the substrate was transferred to another ultrahigh vacuum MBE chamber without breaking vacuum for the growth of the 3D TI \BSTS{} thin film (8 nm). The Dirac fermion dynamics in \BTSx{} can be engineered by varying the composition of Te (3-x) and Se (x)\cite{Chen2013}, and we chose x to be 1 (\BTTS) to place the Dirac point above the top of the valence band. Further engineering of the chemical potential was achieved by Sb-doping: with an optimal ratio of Bi and Sb (Bi:Sb $\approx$ 1.25:0.75) we were able to place the chemical potential into the bulk band gap, as confirmed by the electrical transport measurements.


\begin{figure*}
\includegraphics[width=160mm]{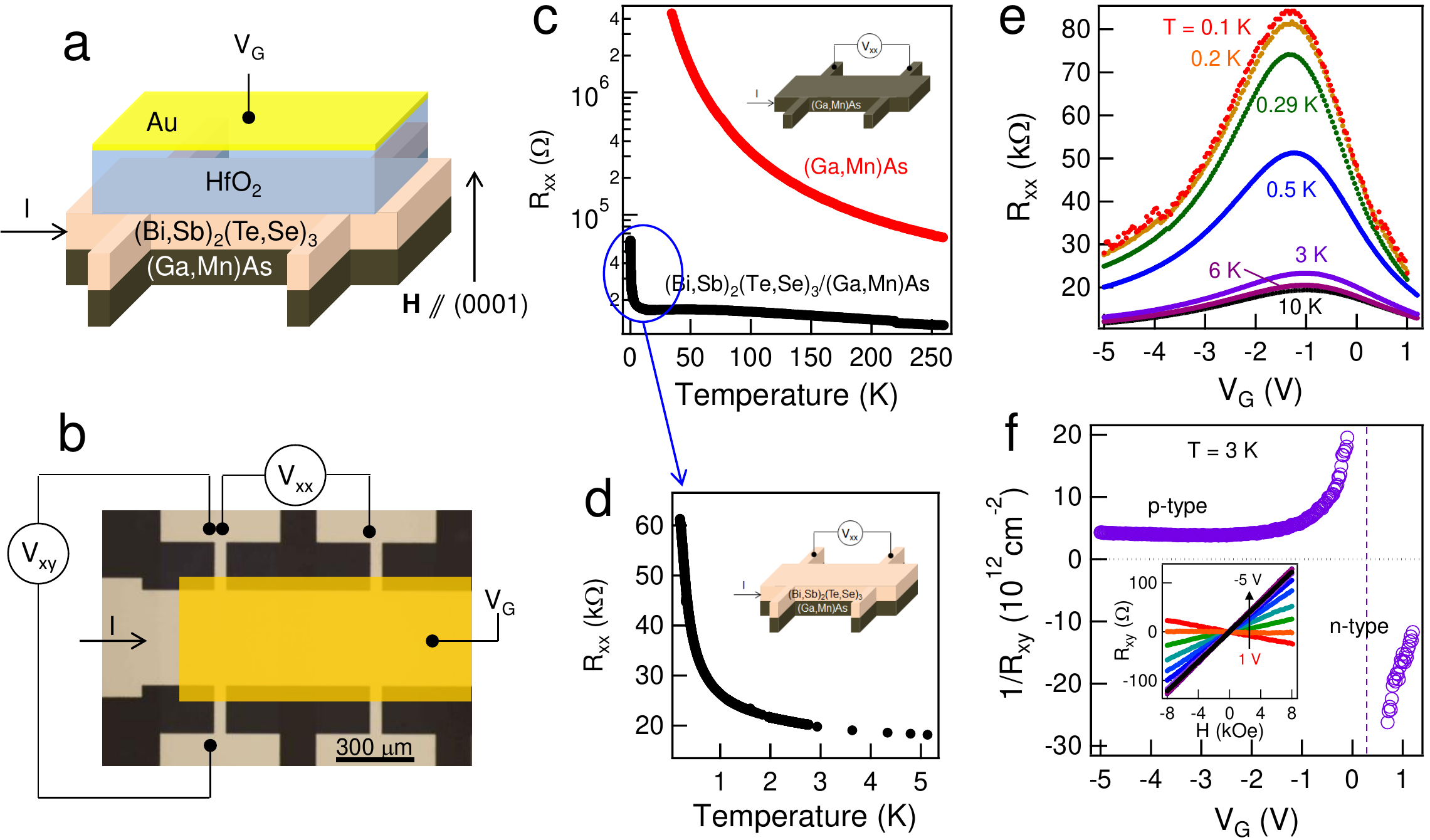} 
\caption{\textbf{Lateral electrical transport properties with temperature and gate-voltage sweep.} \textbf{a}, A schematic of a top-gated TI/FMI Hall-bar device. \textbf{b}, An optical microscope image of a Hall-bar device (650 $\times$ 400 $\mu$m$^2$) with false-colored Au gate metal and the measurement description. \textbf{c}, Temperature dependence of the longitudinal sheet resistance $R_{xx}$ of \BSTS/\GMA{} heterostructure channel (black circles) and of only \GMA{} channel (red circles). Inset is a schematic of device and measurement description of \GMA{} layer. \textbf{d}, $R_{xx}$ of \BSTS/\GMA{} heterostructure in the low temperature regime. Inset is a schematic of device and measurement description of the heterostructure. \textbf{e}, Gate-voltage dependence of the longitudinal sheet resistance $R_{xx}$ with zero magnetic field at different temperatures from 0.1 K (red) to 10 K (black). \textbf{f}, Inverse of Hall resistance $R_{xy}$ in the unit of 2D carrier concentration at 3 K. Inset is the results of Hall measurements with different gate voltages from 1 V to -5 V at 3 K.} 
\label{gating}
\end{figure*}

Although the selective modification of one TI surface with an FMI is advantageous, as discussed earlier, the buried interface between TI and FMI restricts direct probing of the modified TI surface state by techniques such as ARPES or scanning tunneling microscopy. However, electrical transport measurements do provide a route to study the modification of the surface states by quantum corrections to the magneto-conductance (MC) and by the anomalous Hall effect (AHE). For the transport measurements, we fabricated a top-gated Hall-bar device with high-$\kappa$ dielectric HfO$_2$ and Au/Ti gate metal by standard photo-lithography (Figs.\ref{gating}a and \ref{gating}b). One important question for the electrical transport laterally through the heterostructure is whether a current flows only through the TI layer. The black curve in the Figs.\ref{gating}c and \ref{gating}d represents the resistivity when the current flows through the whole TI/FMI heterostructure while the red curve shows the resistivity of only \GMA{} layer after the TI overlayer was carefully removed by mechanical scratching. Since the resistivity of the \GMA{} is more than two orders of magnitude higher than that of the heterostructure below 40 K, we conclude that the current flows mostly through the TI layer in the measurement range. Longitudinal sheet resistance $R_{xx}$ and Hall resistance $R_{xy}$ of the top-gated TI/FMI device under the gate-voltage sweep show the typical ambipolar transport behavior of TI films (Figs. \ref{gating}e and \ref{gating}f). $R_{xy}$ changes its sign at around $V_G =$ 0.3 V while $R_{xx}$ meets maximum at $V_G =$ -1.3 V, indicating that the chemical potential is tuned from a position above the Dirac point to a position below the Dirac point near the top of the valence band. The mismatch of the gate voltages between the charge neutrality point (the sign change of $R_{xy}$) and the $R_{xx}$ peak reveals that the carrier densities of top and bottom surfaces do not match each other and the two surfaces could even have different types of carriers in a certain range of gate voltage. The chemical potential of the surface state at the bottom surface interfaced with the FMI may not be determined solely from the gate-voltage dependence of the channel resistance and the Hall resistance of the whole TI layer, but careful studies of quantum corrections to MC and AHE make it possible to determine the chemical potential of the surface state at the bottom surface interfacing the \GMA{} layer.


\begin{figure}
\includegraphics[width=160mm]{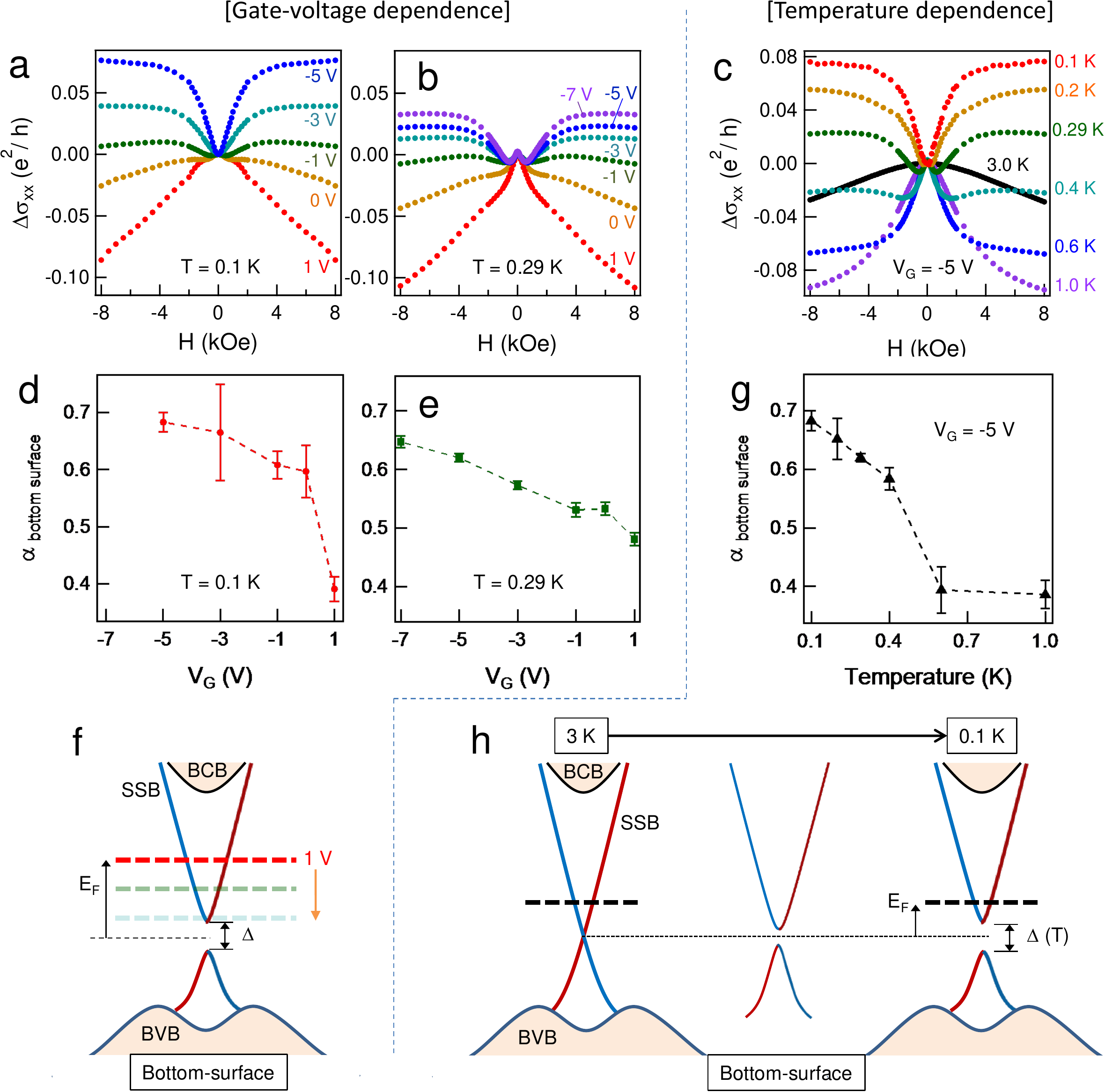} 
\caption{\textbf{Quantum corrections to MC.} \textbf{a,b}, MC with applied gate voltages at 0.1 K and 0.29 K, respectively. Crossover from positive MC (WAL) to negative MC (WL) is observed as gate voltage decreases from 1 V. \textbf{c}, Similar crossover with a temperature range of 3.0 $-$ 0.1 K at $V_G =$ -5 V. \textbf{d,e} Gate-voltage dependence of the prefactor $\alpha_1$ for bottom surface ($\alpha_{bottom}$ $_{surface}$) at 0.10 K (red circles) in \textbf{d} and 0.29 K (green squares) in \textbf{e}, fit by Eq.(\ref{HLN2band}). \textbf{f}, A cartoon of the band diagram of the bottom surface interfaced with \GMA. The size of an opened gap ($\Delta$) is fixed at a certain temperature and the chemical potential is tuned from above and towards the energy gap as the gate voltage decreases from 1 V. \textbf{g}, Temperature dependence of the prefactor $\alpha_{bottom surface}$ with $V_G =$ -5 V (black triangles). \textbf{h}, A cartoon of the band diagram of the bottom surface. The size of the energy gap widens as temperature decreases.}
\label{MC}
\end{figure}

We now discuss the results of magneto-transport measurements from a top-gated Hall-bar device with both tuning gate voltage and varying temperature (Fig. \ref{MC}). In Figs. \ref{MC}a and \ref{MC}b, at fixed temperatures (0.1 K and 0.29 K) MC with $V_G =$ 1 V shows a positive behavior. As the gate voltage decreases, the positive MC gradually changes to negative MC. One salient feature in the crossover regime is the coexistence of positive MC peak near zero magnetic field and negative MC for larger magnetic fields. Interestingly, a similar crossover from positive MC to negative MC is also seen with lowering temperature at a fixed gate voltage (-5 V) as shown in Fig. \ref{MC}c. MC shows positive, parabolic behavior down to 3 K, and as the temperature decreases, the positive MC becomes sharper near zero magnetic field and changes to negative MC. The similarity for the crossovers in MC by gate-voltage dependence as well as by temperature dependence suggest one unified physical picture for both cases. 

The corrections to the classical MC near zero magnetic field can be understood as quantum corrections of weak antilocalization (WAL) and weak localization (WL)\cite{Suzuura2002}. (See supplementary information for discussions on WAL in TI.) 
The FMI is expected to break TR symmetry and open a gap at the Dirac point. The size of the gap ($\Delta$) together with the position of the Fermi energy ($E_F$) modifies the $\pi$-Berry phase of the TI surface state $\varphi=\pi(1-\Delta/2E_F)$ and causes a crossover between WAL and WL\cite{Lu2011}. 
When the gap is closed ($\Delta$ = 0) or $E_F$ is far away from the Dirac point ($E_F \gg \Delta$), the Berry phase remains as $\pi$, leading to WAL. On the other hand, when $E_F$ approaches the gap by tuning the gate voltage or increasing the gap $\Delta$ by lowering temperature, especially in the case when $\Delta = 2E_F$, the Berry phase goes to 0, leading to WL. 
This WAL-WL crossover in MC has been experimentally demonstrated in TI systems with an energy gap opened by either hybridization or magnetic doping/proximity\cite{Lang2013,Zhang2012,Bao2013,Cha2012,Liu2012,zhang2014,Yang2013}.
Here we use a relatively thick TI film with no contribution from hybridization between the surfaces to open the gap as well as demonstrate the crossover by varying both the chemical potential and temperature. Notably, in certain regimes of temperatures and gate voltages, we observe the unique coexistence of a WAL peak near zero magnetic field and WL behaviors at larger magnetic fields, indicating at least two decoupled transport channels with different dephasing lengths. As the gate voltage is tuned from -5 V to 1V at 0.29 K, MC becomes negative (WL) for larger magnetic field while the positive peak (WAL) is still observed for small magnetic field near zero, as shown in Fig. \ref{MC}b. We identify the contribution of the WL transition from the gapped surface state of the bottom surface due to proximity with the FMI. The persisting WAL peak near zero magnetic field comes from the gapless surface state of the top surface decoupled from the bottom surface.  

To quantitatively analyze the WAL and WL, we set up a model for the lateral transport in a TI film of which the bottom surface is interfaced with an FMI. We employ the standard equation for the quantum corrections to MC [Eq. (S1)]. It is important to consider the strength of the coupling between the top and bottom surfaces. In the range of gate voltage (-5 V $-$ 1 V), the chemical potential of TI is mostly in the region where the bulk is depleted and the top and bottom surfaces are decoupled, as indicated by the carrier density ($n_{2D} \cong 4.3 \times 10^{12}$ cm$^{-2}$) at $V_G =$ -5 V which is low enough to place the chemical potential in the bulk band gap while at $V_G =$ 1 V the chemical potential is still in the ambipolar transport region (Fig. \ref{gating}f). 
In the case of no magnetic proximity to a surface of a TI film, fitting the measured MC by the one-band formula gives the prefactor $\alpha$ as -1\cite{Checkelsky2011,Steinberg2011,Chen2011}. Each surface contributes -1/2 to the resulting $\alpha=-1$ for both surfaces. In the case of the magnetic proximity to a surface of the TI film, modification of one surface would not affect the resulting contribution of $\alpha=-1/2$ from the other surface when two surfaces are electrically decoupled. This is the case for our TI system in which one surface state is modified by an adjacent FMI. Although we focus on the range of gate voltage in which the top and bottom surfaces are decoupled, we do not observe $\alpha$ approaching -1 by one-band fitting. Thus, we use a two-carrier model for decoupled top and bottom surfaces with the first prefactor $\alpha_0$, for a gapless surface state, fixed to -1/2:
\begin{equation}
\Delta\sigma(H)=\sum\limits_{i=0,1} \alpha_i\frac{e^2}{2\pi^2\hbar}\left[\psi\left(\frac{1}{2}+\frac{\hbar c}{4el^2_{\phi,i} H}\right)-\ln \left(\frac{\hbar c}{4el^2_{\phi,i} H}\right)\right],\:\alpha_0 = -1/2.
\label{HLN2band}
\end{equation}

The quantitative results of $\alpha_1$ of the bottom surface provides an indirect way to estimate the $E_F$ position of the bottom surface. At a given temperature, the size of a magnetic gap does not change but $E_F$ is tunable by electrical gating. Figures \ref{MC}d and \ref{MC}e show that $\alpha_{1}$ increases as the gate voltage decreases from 1 V to -5 V (-7 V) at 0.1 K (0.29 K). This can be interpreted as $E_F$ at the bottom surface being tuned from above towards the magnetic gap, but not passing through it, as illustrated in Fig. \ref{MC}f. This qualitatively agrees with the WAL-WL crossover when the Berry phase changes from $\pi$ to a smaller value by tuning $E_F$. Similarly, as we fix the gate voltage to -5 V to place $E_F$ close to the gap and vary temperature, $\alpha_{1}$ increases with decreasing temperature (Fig. \ref{MC}g) as $\Delta$ increases at lower temperatures due to the temperature dependence of the interfacial exchange coupling with the adjacent \GMA{} layer (illustrated in Fig. \ref{MC}h). 


\begin{figure*}
\includegraphics[width=90mm]{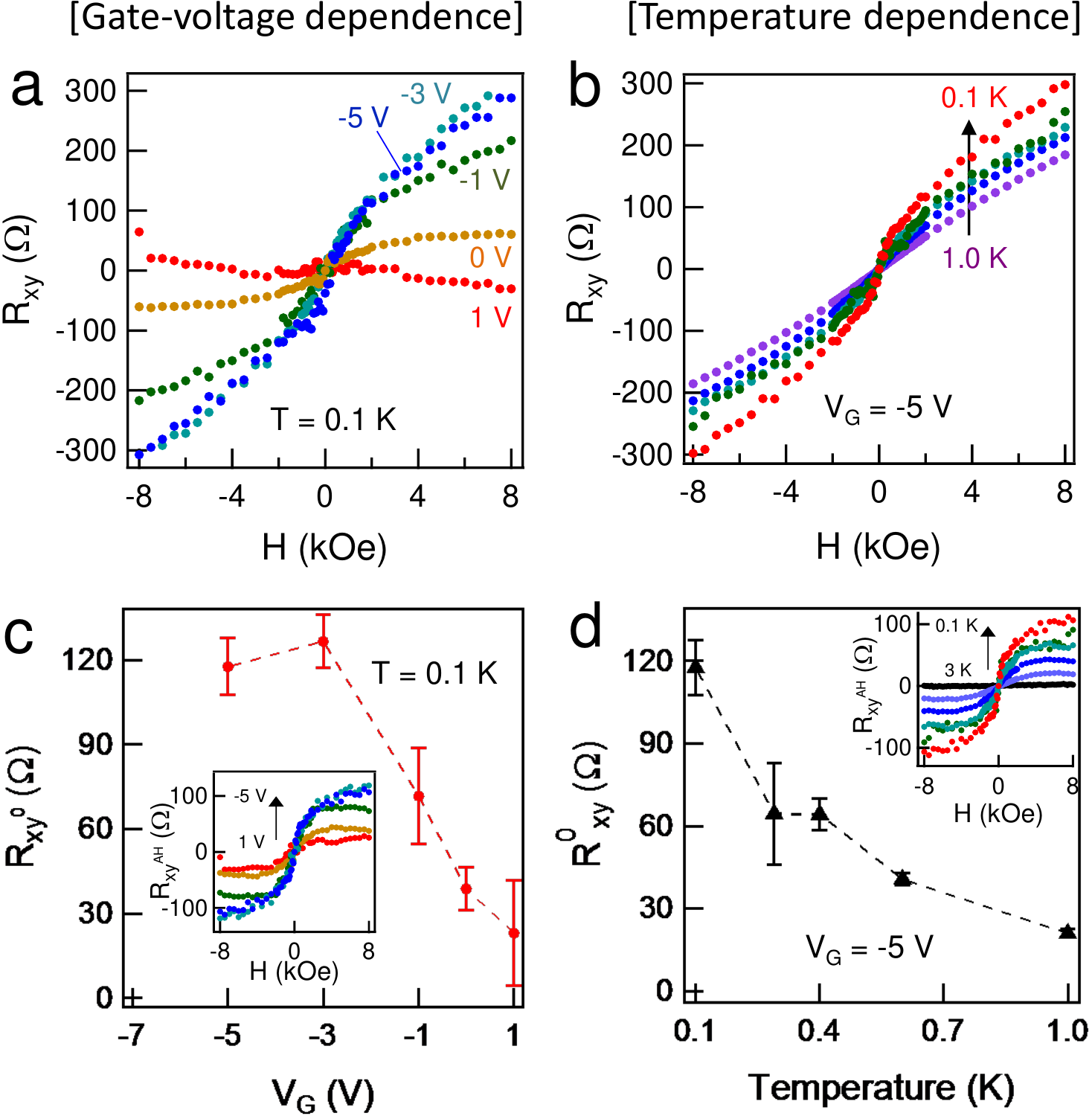} 
\caption{\textbf{Anomalous Hall effect.} \textbf{a}, Hall resistance $R_{xy}$ vs magnetic field with applied gate voltages at 0.1 K. The sign change of the slope indicates n- to p-type carrier change by applying gate voltage. \textbf{b}, Hall resistance $R_{xy}$ vs magnetic field below 1.0 K. Clearer AHE is seen as temperature decreases. \textbf{c}, Gate-voltage dependence of the magnitude of the anomalous Hall resistance $R_{xy}^{0}$ at 0.1 K (red circles). Inset is the anomalous Hall resistance $R_{xy}^{AH}$ at each $V_G$: 1 V (red), 0 V (orange), -1 V (green), -3 V (cyan), and -5 V (blue). \textbf{d}, Temperature dependence of $R_{xy}^{0}$ with $V_G =$ -5 V. Inset is the anomalous Hall resistance $R_{xy}^{AH}$ at 3.0 K (black), 1.0 K (violet), 0.6 K (blue), 0.4 K (cyan), 0.29 K (green), and 0.1 K (red). 
} 
\label{AHE}
\end{figure*}

Now we discuss the results from Hall measurements (Figs. \ref{AHE}a and \ref{AHE}b) where we observe a systematic emergence of AHE as the temperature decreases or as the chemical potential is lowered by tuning gate voltages. The absence of hysteresis of the AHE indicates that long-range magnetic ordering is not induced in the bottom surface of the TI film by the adjacent \GMA{} layer in the measurement range of temperature (100 mK $-$ 10 K) and gate voltage (-7 V $-$ 1 V).  
To understand the anomalous Hall contribution from the interface between TI and FMI, we consider a 2D Dirac model with a finite energy gap ($\Delta$). A direct calculation of the Hall conductance leads to
\begin{equation}
\sigma_{xy}(-\frac{|\Delta|}{2}<E_F<\frac{|\Delta|}{2}) = \frac{e^2}{2h},   \sigma_{xy}(E_F>\frac{|\Delta|}{2}) = \frac{e^2}{2h} \frac{\Delta}{2E_F}
\label{sigmaxytot}
\end{equation}
with $E_F= \sqrt{(\hbar v_F k)^2 + (\Delta/2)^2}$ (See supplementary information for detailed calculation). Equation (\ref{sigmaxytot}) shows that the Hall conductivity is half-quantized in the insulating regime of a single Dirac model, and the half-integer quantum Hall conductivity monotonically decreases as $E_F$ moves above the energy gap or as the gap gradually closes with a $E_F$ fixed to a position near the gap. 

\begin{figure*}
\includegraphics[width=90mm]{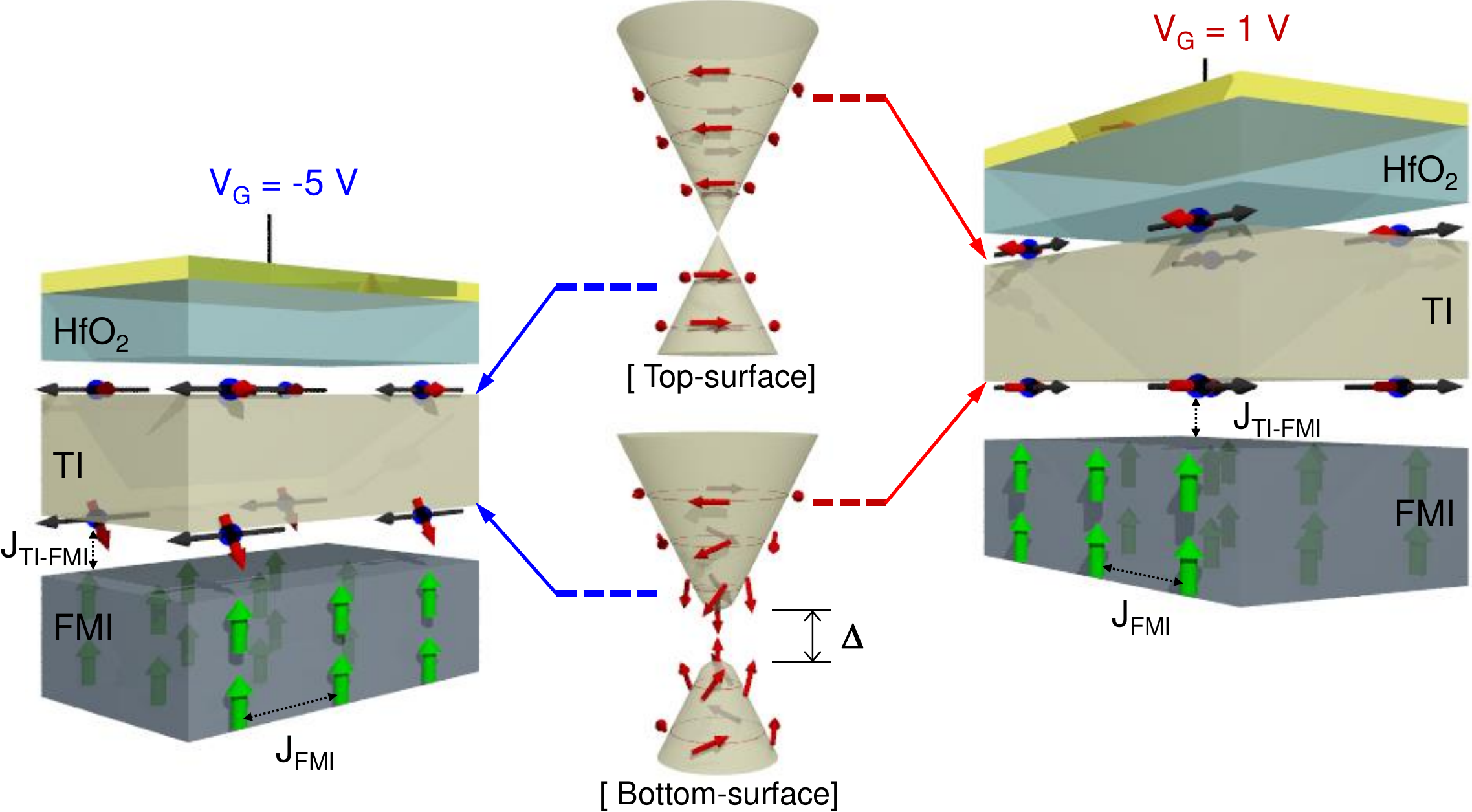} 
\caption{\textbf{Spin textures in TI surfaces.} An illustration of a gapless Dirac surface state for top surface, a magnetic-gapped surface state with hedgehog spin texture for bottom surface, and top-gated TI/FMI heterostructure devices with $V_G =$ -5 and 1 V. Green arrows represent the magnetization of the FMI \GMA{} layer and gray arrows and red arrows in TI layer represent the directions of electron propagation and its spin, respectively. $J_{FMI}$ is the exchange-coupling constant between Mn moments in \GMA{} while $J_{TI-FMI}$ is the relatively weaker exchange-coupling constant between electrons in TI bottom surface and Mn moments in \GMA.}
\label{Illustration}
\end{figure*}

Since our results are not in the regime of the quantized Hall conductivity, the observed Hall conductivity is smaller than $e^2/2h$. However, it follows the qualitative behavior of Eq. (\ref{sigmaxytot}). Figures \ref{AHE}c(inset) and \ref{AHE}d(inset) clearly show the systematic emergence of the anomalous Hall term $R_{xy}^{AH}$ with respect to the gate voltage and temperature, where $R_{xy}^{AH}$ is obtained after subtracting the ordinary Hall term $R_{xy}^{OH}$ from the Hall resistance as $R_{xy} ^{AH}=R_{xy} - R_{xy}^{OH}$. 
We show $R_{xy}^{AH}$ instead of $\sigma_{xy}^{AH}$ since $R_{xx}$ term contains a large contribution of the top surface and affects the values of $\sigma_{xy}=R_{xy}/(R_{xx}^2+R_{xy}^2)$. The expression for $R_{xy}$ from $\sigma_{xy}$ can be written as:
\begin{equation}
R_{xy}(E_F>\frac{|\Delta|}{2})=\frac{\sigma_{xy}}{\sigma_{xx}^2+\sigma_{xy}^2} \cong \frac{m^2}{2he^2\tau^2}\frac{1}{n^2}\frac{\Delta}{2E_F}
\label{Rxy}
\end{equation}
with $\sigma_{xx}$ from the Drude model $\sigma_{xx}=e^2\tau n/m$ where \textit{n}, \textit{m} and $\tau$ are the carrier density, the effective mass, and the relaxation time between collisions. 
For $E_F>|\Delta|/2$, \textit{n} increases as $E_F$ increases (moves away from the gap). 
Similarly to the case of quantum corrections to MC, the change of $R_{xy}$ reveals a systematic modification of the size of the energy gap and the chemical potential. As the chemical potential lowers and approaches the energy gap by tuning the gate voltage from 1 V to -5 V, the estimated magnitude of the anomalous Hall resistance $R_{xy}^0$ increases. $R_{xy}^0$ is the intercept obtained by extrapolating a linear line of the high field Hall resistance. $R_{xy}^0$ is zero in the case of a closed gap ($\Delta=0$). When a gap opens and widens, a non-zero $R_{xy}^0$ monotonically increases. Figure \ref{AHE}c shows the evolution of the anomalous Hall resistance $R_{xy}^0$ as the chemical potential lowers towards the energy gap. Similarly, Fig. \ref{AHE}d shows the monotonic increase of $R_{xy}^0$ with decreasing temperature, interpreted as the widening of the gap with decreasing temperature. The interpretation of both gate-voltage dependence and temperature dependence of the AHE is consistent with that of the quantum corrections to the MC with varying gate voltage and temperature. The onset temperature of both AHE and WL is much lower than the \TC{} of the adjacent \GMA{} layer, indicating that the exchange coupling between electrons in TI bottom surface and Mn moments in \GMA{} is much weaker than the exchange coupling between Mn moments in \GMA{} (Fig. \ref{Illustration}).

In summary, we synthesized and characterized a TI/FMI heterostructure of a TI film \BSTS{} on a dilute magnetic semiconductor \GMA. The \GMnA{} layer is highly resistive with a perpendicular component of magnetization below 50 K. With an optimal Bi to Sb ratio, the chemical potential was placed in the surface state and further tuned by electrical top gating. The crossover between WAL and WL as well as the systematic emergence of AHE was observed with varying temperatures and gate voltages, interpreted as a result of a gap opening in the Dirac surface state due to the TR symmetry breaking by the exchange coupling between the TI surface state and the adjacent \GMA. The results suggest that the systematic changes in MC and AHE can be used as indirect probes to estimate the $E_F$ position and the existence of the magnetic gap opening for the surface state . 

\bigskip
\bigskip
\textbf{Acknowledgement}
\bigskip

This work was supported in part by DARPA and C-SPIN, one of six centers of STARnet, a Semiconductor Research Corporation program, sponsored by MARCO and DARPA. NS and AR acknowledge additional support from ONR  grant N00014-15-1-2370 and CXL from ONR grant N00014-15-1-2675.

\newpage

\newpage



\section*{Supplementary material}

# Supplementary Information

# Engineering the breaking of time-reversal symmetry in gate-tunable hybrid ferromagnet/topological insulator heterostructures

Joon Sue Lee[*], Anthony Richardella, Robert D. Fraleigh, Chao-xing Liu, Weiwei Zhao, Nitin Samarth[†]

Department of Physics, The Pennsylvania State University, University Park, PA 16802

[*]Current address: California NanoSystems Institute, University of California, Santa Barbara, CA 93106; Electronic address: joonsuelee@engineering.ucsb.edu
[†]Electronic address: nsamarth@psu.edu

**S1. Characterization of $(Bi,Sb)_2(Te,Se)_3$/GaMnAs heterostructure**

Magnetic and structural properties of the $(Bi,Sb)_2(Te,Se)_3$/$Ga_{1-x}Mn_xAs$ heterostructure were studied. The superconducting quantum interference device (SQUID) magnetometry on $Ga_{0.95}Mn_{0.05}As$ with a topological insulator (TI) film on top reveals $T_C \approx 50$ K with a magnetic field perpendicular-to-plane, as shown in Fig.S1a. The elemental composition was determined by energy dispersive spectroscopy within 20% error. Clear hysteresis was seen with a perpendicular magnetic field below the $T_C$ (Fig.S1b). The resulting crystal structure of $Ga_{0.95}Mn_{0.05}As$ film on InP(111)A substrate is GaAs(111) structure. In Fig.1Sc, the x-ray diffraction (XRD) of the $Ga_{0.95}Mn_{0.05}As$ film with a $(Bi,Sb)_2(Te,Se)_3$ film grown on top shows a peak close to the GaAs(111) line. The offset of the $Ga_{0.95}Mn_{0.05}As$ peak from the GaAs(111) line is due to the modified GaAs lattice constant by Mn substitutions and strain of the layer.

An advantage of (Ga,Mn)As as an FM for a TI/FM heterostructure is the well-defined interface without any amorphous growth or secondary phases, as demonstrated in the previous work of epitaxial growth of Bi-chalcogenide TIs on GaAs(111) crystal [1]. The clean interfaces between adjacent layers are shown in the high-resolution transmission electron microscopy (TEM) image and energy-dispersive spectrometer (EDS) scanning transmission electron microscopy (STEM) image in Figures S2 and S3.

## S2. Weak antilocalization in three-dimensional TI

In three-dimensional (3D) TIs where spin and momentum are strongly "locked", weak antilocalization (WAL) naturally arises from the $\pi$ Berry phase of electrons going around the Fermi circle of the Dirac surface state [2]. The quantum corrections to the MC for the Dirac surface states are expected to follow the behavior of the equation for a conventional 2D metal with a strong spin-orbit coupling [3]:

$$\Delta\sigma(H) = \alpha \frac{e^2}{2\pi^2\hbar}[\psi\left(\frac{1}{2} + \frac{\hbar c}{4el_\phi^2 H}\right) - \ln\left(\frac{\hbar c}{4el_\phi^2 H}\right)] \qquad (S1)$$

where $\alpha$ is a prefactor, $\psi$ is the digamma function, and $l_\phi$ is the coherence length. WAL leads to $\alpha$ = -1/2 (symplectic case) while WL by constructive interference leads to $\alpha$ = 1 (orthogonal case). For an ideal 3D TI, when the Fermi energy $E_F$ lies in the bulk conduction band (BCB) or the bulk valence band (BVB), the top and bottom surfaces are coupled by the conducting bulk, and one can expect the prefactor $\alpha$ = -1/2, while $\alpha$ becomes -1 when $E_F$ moves into the bulk band gap where the top and bottom surfaces are decoupled by the insulating bulk [4,5]. We note that the exact value of $\alpha$ can depend on inter-channel scattering [4].

## S3. Hall conductance with a finite energy gap

Let's consider the anomalous Hall contribution from 2D Dirac model. From a theoretical calculation of TI with a single Dirac surface state with a finite energy gap (Δ) at the Γ point in the Brillouin zone, the Hall conductivity is half-quantized as $\sigma_{xy} = e^2/2h$ in the insulating regime ($-\frac{|\Delta|}{2} \leq E_F \leq \frac{|\Delta|}{2}$), with the Hamiltonian written as:

$$H(k) = \hbar v_F(k_x \sigma_y - k_y \sigma_x) + \frac{\Delta}{2}\sigma_z, \qquad (S2)$$

where $\sigma_{x,y,z}$ are the Pauli matrices with basis states of spin-up and spin-down states of real spin, and $v_F$ is the Fermi velocity. The Hall conductance can be evaluated from the TKNN formula [6], and for the above two-band model, a simplified version is given by

$$\sigma_{xy} = \frac{e^2}{h} \int \frac{d^2k}{4\pi} \hat{\mathbf{d}} \cdot \left(\frac{\partial \hat{\mathbf{d}}}{\partial k_x} \times \frac{\partial \hat{\mathbf{d}}}{\partial k_y}\right) \qquad (S3)$$

with the **d** vector defined as $d_x = -\hbar v_F k_y$, $d_y = \hbar v_F k_x$, $d_z = \Delta/2$, $d^2 = (\hbar v_F)^2 k^2 + (\Delta/2)^2$, and $\hat{\mathbf{d}} = \mathbf{d}/|\mathbf{d}|$. Calculation of Equation (S3) using the given **d** vector for the above two-band model gives

$$\sigma_{xy}\left(-\frac{|\Delta|}{2} < E_F < \frac{|\Delta|}{2}\right) = \frac{e^2}{2h} \qquad (S4)$$

in the insulating regime, in which all the contributions of Hall conductivity come from the valence band.

For the metallic regime where $E_F$ lies above the gap ($E_F > |\Delta|/2$), we have additional contribution to Hall conductivity from the conduction band besides the contribution from the valence band, which is given by

$$\sigma_{xy}^c = -\frac{e^2}{h} \int_0^{k_F} \frac{d^2k}{4\pi |\mathbf{d}|^3} \mathbf{d} \cdot \left(\frac{\partial \mathbf{d}}{\partial k_x} \times \frac{\partial \mathbf{d}}{\partial k_y}\right). \qquad (S5)$$

The minus sign accounts for the opposite contribution from the conduction band. Calculation with above **d** vector makes the Hall conductivity as

$$\sigma_{xy}^c = \frac{e^2}{2h}\left[\frac{\Delta}{2}\left((\hbar v_F)^2 k_F^2 + \left(\frac{\Delta}{2}\right)^2\right)^{-1/2} - 1\right]. \tag{S6}$$

Therefore, the total Hall conductivity for the metallic regime is

$$\sigma_{xy}\left(E_F > \frac{|\Delta|}{2}\right) = \frac{e^2}{2h}\frac{\Delta}{2E_F} \tag{S7}$$

with $E_F = \sqrt{(\hbar v_F)^2 k_F^2 + \left(\frac{\Delta}{2}\right)^2}$.

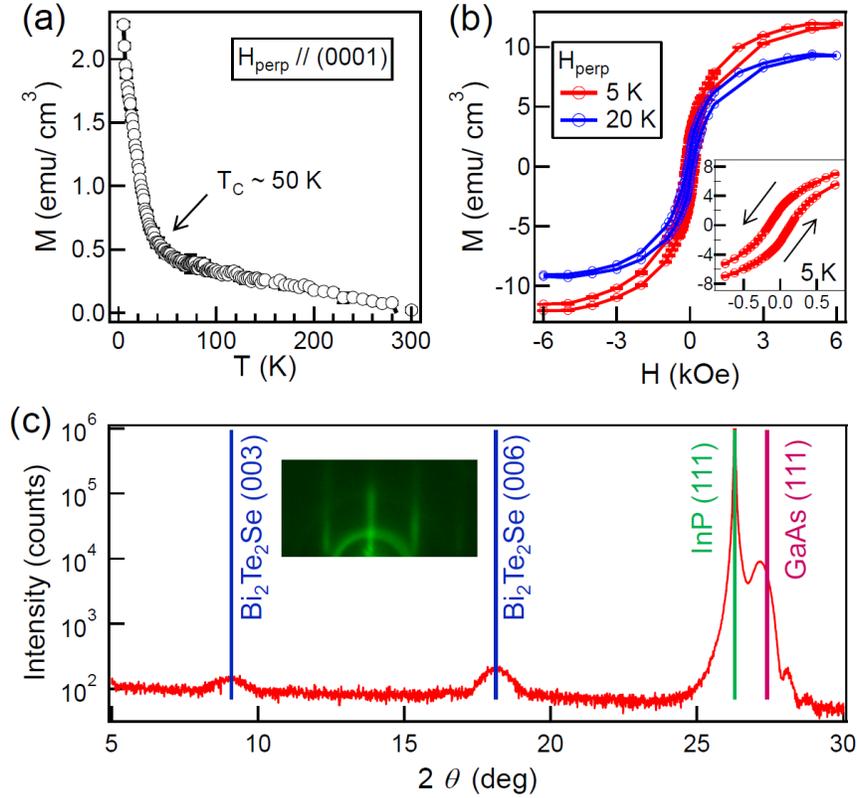

Figure S1. **a**, Temperature dependence of magnetization with perpendicular-to-plane magnetic field for the $(Bi,Sb)_2(Te,Se)_3/Ga_{0.95}Mn_{0.05}As$ heterostructure by SQUID magnetometry. $T_C$ of the (Ga,Mn)As layer is around 50 K. **b**, Magnetization of the (Ga,Mn)As layer as a function of perpendicular-to-plane magnetic field below $T_C$. Clear hysteresis was observed as shown in the inset. **c**, XRD of the $(Bi,Sb)_2(Te,Se)_3$/(Ga,Mn)As heterostructure on InP (111)A substrate. Blue, green, and violet lines are the calculated positions of $Bi_2Te_2Se$ (003), $Bi_2Te_2Se$ (006), InP (111), and GaAs (111) peaks, respectively. The inset shows the typical reflection high-energy electron diffraction patterns observed while the $(Bi,Sb)_2(Te,Se)_3$ film growth.

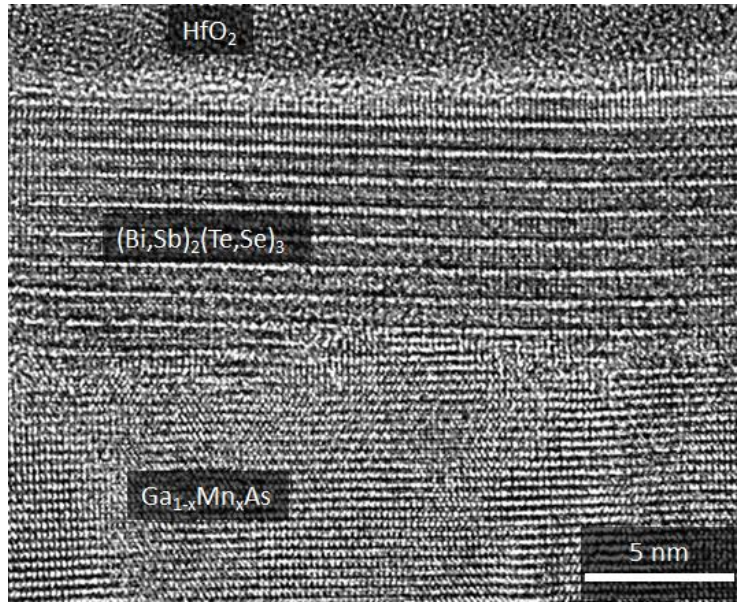

Figure S2. A TEM image of $(Bi,Sb)_2(Te,Se)_3$/(Ga,Mn)As heterostructure with a $HfO_2$ layer on the TI film. An interface with no amorphous layer is seen between $(Bi,Sb)_2(Te,Se)_3$ and (Ga,Mn)As.

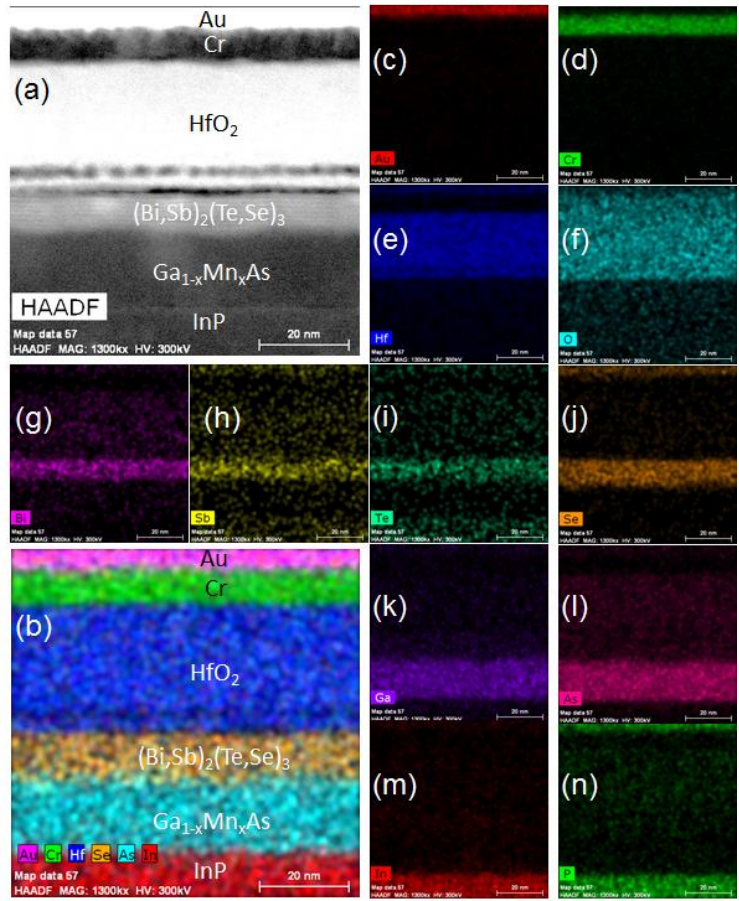

Figure S3 (a) High-angle annular dark field (HAADF) STEM image of an Au/HfO$_2$ top-gated (Bi,Sb)$_2$(Te,Se)$_3$/(Ga,Mn)As heterostructure. (b) EDS mapping image of Au, Cr, Hf, Se, As, and In, which represent each layer of the top-gated heterostructure device. (c-n) EDS mapping for each element of (c) Au, (d) Cr, (e) Hf, (f) O, (g) Bi, (h) Sb, (i) Te, (j) Se, (k) Ga, (l) As, (m) In, and (n) P.



\end{document}